# Crystal structures of superconducting phases of S and Se

Olga Degtyareva[1], Eugene Gregoryanz[1], Maddury Somayazulu[2],
Ho-kwang Mao[1], and Russell J. Hemley[1]

[1] *Geophysical Laboratory, Carnegie Institution of Washington,
5251 Broad Branch Rd. N.W., Washington D.C. 20015, USA.*
[2] *HPCAT, Carnegie Institution of Washington, APS,
9700 South Cass Avenue, Argonne, IL 60439, USA.*

Compressed S and Se are studied by x-ray diffraction with synchrotron radiation up to 160 GPa. The S-IV phase is shown to have a body-centered monoclinic structure and to be stable between 83 and 150 GPa on pressure increase, where it transform to S-V with a primitive rhombohedral β-Po structure. Observation of the modulation reflections in S-IV up to 135 GPa shows that its crystal structure is incommensurately modulated, as recently reported for Se and Te. The modulated body centered monoclinic phase of Se, Se-IV, is shown to transform to the β-Po phase Se-V at around 80 GPa. Se-V transforms to a body-centered cubic phase at 140 GPa in accordance with previous studies. Pressure dependence of the structural parameters of these high-pressure phases in S and Se are discussed in relation to their superconducting behaviour.

## I. Introduction

Recent discoveries of incommensurate host-guest structures[1–4] and incommensurately modulated structures[5–7] in elemental metals at high pressure suggest that aperiodic structures are a common phenomenon among the elements under pressure. The previously developed formalism of superspace crystallography[8] has been successfully applied to the description of aperiodic structures of the elements.[5–7,9,10]

Aperiodic structure with large modulations in atomic positions is found in metallic phases of group-VI elements Te and Se, Te-III and Se-IV,[6,7] and is described as a body-centered monoclinic (*bcm*) cell with two atoms in the unit cell and a modulation wave running along the *b*-axis. Atoms are displaced from their ideal positions according to this harmonic wave mainly along the *c*-axis. This *bcm* structure can be considered to be a distortion of the β-Po (primitive rhombohedral) structure. On compression, the monoclinic distortion of Te-III and Se-IV decreases and the *bcm* structure approaches that of β-Po; at the same time the modulation reflections become weaker indicating a decrease of the modulation amplitude. To distinguish between the modulated monoclinic and the rhombohedral β-Po structures, one needs to obtain high-resolution diffraction data with low background and high signal-to-noise ratio. Such experimental studies[6] show that Te-III with a modulated *bcm* structure, stable above 4.5(2) GPa, transforms directly to a *bcc* structure (Te-V, Ref. 11) at 29.2(7) GPa without reaching the β-Po structure, while Se-IV with a modulated *bcm* structure is shown to be stable from 28 GPa to at least 70 GPa.[7] Assuming a continuous phase transition from a modulated *bcm* to β-Po structure, extrapolation of the Se-IV structural parameters suggested a transition pressure of 82 GPa,[7] while the β-Po phase of Se is known to transform to *bcc* at 140

GPa.[12] These results[6,7] required a revision of the phase transition sequence of the metallic Te and Se reported earlier based on the lower-resolution diffraction data.[12–18] However, the question whether the modulated Se-IV phase transforms to $\beta$-Po above 70 GPa remains open and the nature of this transition (first or second order) remains unclear.

The lighter group-VI element S with its metallic phases has been reported to be isostructural with Te and Se.[19,20] S-IV is stable above 83 GPa with a base-centered orthorhombic (*bco*) structure,[19,20] the same as previously found in Te-III and Se-IV. S-IV is reported to transform to the rhombohedral $\beta$-Po structure at 162(5) GPa, which has been shown to be stable up to at least 212 GPa.[20] The stability of the *bco* structure for sulfur is not confirmed by theoretical calculations,[21,22] similar to the calculations on Se[23] and Te[24] that did not reproduce the stability of the *bco* structure. Thus, in light of recent experimental reports of incommensurately modulated crystal structures of Te-III and Se-IV,[6,7] the crystal structure of S-IV should be re-examined with the advanced experimental techniques. Moreover, the metallic phases of S, Se and Te discussed above are known to be superconductors showing complex pressure dependence of superconducting transition temperature (Ref. 25 and references therein). Precise determination of the crystal structure of metallic sulfur and selenium may help to understand the complex behavior of $Tc$ in the group-VI elements.

Here, we report the determination of the crystal structure of S-IV, stable above 83 GPa, as an incommensurately modulated. We observe the structural modulation in S-IV up to 135 GPa, the highest pressure a modulated structure has been reported so far. Also, we show that the phase transition to $\beta$-Po in S occurs at a lower pressure of 153 GPa than previously thought. Superconducting properties and pressure dependence of $Tc$ in S and Se are discussed in light of present structural data.

## II. Experimental Methods

Samples of 99.999% purity of S (Puratronic) and of 99.99% purity of Se (Alfa Aesar) were loaded in Mao-Bell diamond anvil cells (DAC's) with an opening allowing probing up to 27 degrees of $2\theta$ of the angle-dispersive x-ray diffraction. Powder x-ray diffraction data were collected at beam line 16-ID-B (HPCAT) at the Advanced Photon Source. Focused monochromatic beams of different wavelengths ($\lambda$ = 0.36-0.42 Å) were used and the data were recorded on MAR image plate calibrated with a CeO2 or Si standard. Samples of S and Se were studied at room temperature up to pressures of 160 and 150 GPa, respectively. Samples were loaded in DAC's with neon as a pressure transmitting medium in order to obtain the diffraction patterns of high-pressure phases with less stress and strain. Our diffraction patterns show that the starting material was always the known ambient orthorhombic S-I and trigonal Se-I phases. To determine the pressure, we used *in situ* fluorescence measurements of ruby chips that were loaded in the sample chamber. The well resolved ruby spectra show that neon provides quasi-hydrostatic condition. Powder diffraction data were integrated azimuthally using FIT2D,[26] and structural information was obtained by full structure Rietveld refinement of the integrated profiles of S and Se, respectively, using JANA2000.[27]

## III. Results

With an increase in pressure above 36 GPa, we obtain diffraction patterns that correspond to the S-III phase, reported recently as having body-centered tetragonal structure with 16 atoms in the unit cell, space group $I4_1/acd$.[28,29] Our present data give the following lattice parameters for S-III at 76 GPa: $a$ = 7.706(1) Å, $c$ = 3.061(1) Å, atomic volume $Vat$ = 11.36 Å$^3$. Diffraction lines of S-IV phase appear at 83 GPa observed together with the remaining S-III, and the diffraction patterns of pure S-IV are observed above 94 GPa. We obtained a quasi-single crystal of S-IV, which gave several arcs on a diffraction pattern (Fig. 1, upper panel). At each pressure, three patterns were collected from the sample placed at 0 degree and ±4.0 degree between the DAC axis and the incident beam, to obtain information from a larger portion of the reciprocal space. The integrated profile is obtained from integration of the diffraction arcs. We extracted reliable information on peak positions; the obtained intensities, however, cannot be used in the analysis of atomic positions. The resulting integrated diffraction pattern obtained from S-IV at 102(3) GPa is shown in Fig. 1 (lower panel). The fitting of this pattern with the previously proposed *bco* structure19 (the predicted peak positions from the *bco* structure correspond to the upper tick marks of the *bcm* structure in Fig. 1) shows that some reflections remain unfitted. These extra reflections are similar in their positions to the positions of the modulation reflections reported in literature for Te and Se in their modulated phases Te-III and Se-IV[6,7] and in our own diffraction data on Se-IV (Fig. 2). Application of the incommensurately modulated structure model, proposed for Te-III[6] to the diffraction pattern of S-IV at 102(3) GPa shows that the positions of the extra reflections are fitted as modulation reflections (lower tick marks in Fig. 1, lower panel).

The structure of S-IV is body-centered monoclinic with two atoms in the unit cell (Figs. 3a and b) with the lattice parameters (obtained from the least-square fit of the diffraction peak positions) $a$ = 2.8217(8) Å, $b$ = 3.4698(9) Å, $c$ = 2.2187(4) Å, $\beta$ = 112.95(3)$^o$, atomic volume V$at$ = 10.00(1) Å$^3$ at 102(3) GPa, and the superspace group $I'2/m(0q0)s0$, where $I'$ denotes centering in 4D. As was proposed for Te-III and Se-IV,[6,7] the atoms in S-IV are displaced from their ideal positions (0 0 0) and (1/2 1/2 1/2) in the $x$ and $z$ direction due to a harmonic modulation wave running along the $b$-axis. The amplitude of the displacement of the atoms due to the modulation wave in S-IV could not be determined in present work because of the insufficient information on the diffraction intensity. However, the reliably determined peak positions of the modulation reflections allow us to determine the period of the modulation wave in S-IV, which is incommensurate with the lattice period $b$ resulting in modulation vector $(0q0)=(0,0.289(1),0)$ at 102(3) GPa.

Using this structural model, all the observed diffraction reflections can be indexed (Fig. 1) using four integers $h,k,l,m$, according to $\mathbf{H} = h\mathbf{a}^* + k\mathbf{b}^* + l\mathbf{c}^* + mq\mathbf{b}^*$, where $\mathbf{a}^*,\mathbf{b}^*,\mathbf{c}^*$ define the reciprocal lattice of the *bcm* structure. Several modulation reflections (with mod($m$) = 1) are clearly visible in the diffraction image of S-IV at 102(3) GPa (Fig. 1). These modulation reflections, though with decreased intensity, are still visible in the 129 GPa diffraction pattern of S-IV (Fig. 2a). Only a faint trace of one modulation reflection is found in the 135 GPa diffraction image, and in the 146 GPa image the modulation reflections are completely absent. Thus, similar
to the behavior reported for Te-III and Se-IV,[6,7] the intensity of the modulation reflections in S-IV decreases with pressure. The pressure dependence of the value of the modulation vector $q$ for S-IV is obtained from the least-square fit of the 5-10 (main and modulation) diffraction reflections using the U-fit computer program[30], and the result is shown in the inset to Fig. 4. The value of the modulation vector $q$ decreases in S-IV from 0.289(1) at 102(3) GPa down to 0.269(1) at 129 (3) GPa.

With increasing pressure, we observe the doublet splitting of the main reflections 1100 and -1010 (at $2\theta$ around 10.2 degree) and reflections 0110 and 0200 (at $2\theta$ around 12.1 degree) (Fig. 1 and 2a) to decrease in the pressure range from 100 to 150 GPa and to become a singlet within the error at 156 GPa. The pattern of S at 160 GPa (Fig. 2b) is fitted with primitive rhombohedral $\beta$-Po structure, with $a_h$ = 3.3780(1) Å, $c_h$ = 2.6919(3) Å (in hexagonal setting), and $a_r$ = 2.147(1) Å, $\alpha_r$ = 103.77(1)$^0$ (in rhombohedral setting), and resulting atomic volume V$at$ = 8.87 Å$^3$. Thus, the *bcm* structure of S-IV transforms to $\beta$-Po structure of S-V between 150(3) and 156(3) GPa, and the transition pressure is estimated as 153(5) GPa. This transition pressure is lower than the pressure of 162(5) GPa reported in Ref. 20, which had a low resolution data that could lead to a big uncertainty in determination of a structural distortion. In contrast to Ref. 20, our present data allowed us to observe the small splitting of the reflections, that determine the monoclinic distortion of the $\beta$-Po structure, because of the spatial separation of these reflections on the diffraction image (Fig. 1).[31]

We also performed high-pressure studies of the metallic phases of the next member of the chalcogen family, Se. On pressure increase above 28 GPa, we obtain the Se-IV phase with an incommensurately modulated *bcm* structure, in accordance with previous study[6,7]. Our Rietveld refinement of Se-IV at 46(1) GPa (Fig. 2c) gives the following structural parameters: $a$ = 3.2226(1) Å, $b$ = 3.9896(1) Å, $c$ = 2.5596(1) Å, $\beta$ = 113.265(4)$^o$ and V$at$ = 15.12(1) Å$^3$. The incommensurate wave vector at the same pressure is (0,0.2847(6),0). The value of the modulation vector $q$ is observed to decrease with pressure in the pressure range between 35 and 53 GPa, (Fig. 4, insert in the middle panel), in accordance with Ref.7. The refined modulation parameters, which characterize the displacements of atoms from their ideal positions due to the modulation wave, are $B1x$ = 0.020(2) and $B1z$ = 0.083(1) at 46(1) GPa. These values are in agreement with those reported earlier for Se-IV.[7] In present work, we were able to observe the modulation reflections only up to pressure of 53 GPa, while the study in Ref. 7 reports the observation of the modulation reflections up to the maximum pressure of 70 GPa reached in that study. The difference in observations can be possibly accounted for by the different pressure transmitting media used in our study and Ref. 7.

On pressure increase from 28 GPa to 70 GPa, the splitting of the main reflections in Se-IV decreases, and, as mentioned above, so does the intensity of the modulation reflections, indicating a reduction of the monoclinic distortion. Modulation reflections can be used to obtain accurate lattice parameters of Se-IV and to determine the degree of the monoclinic distortion, as shown in Ref. 7. For the diffraction patterns of Se obtained in present work in the pressure range between 70 and 90 GPa, it is impossible to distinguish between the *bcm* and $\beta$-Po structures, and although the diffraction patterns resemble those that would be expected from a primitive rhombohedral phase, we cannot rule out a slight monoclinic distortion in this pressure range. At 90 GPa, the observed diffraction pattern can be fitted with primitive rhombohedral $\beta$-Po structure, indicating that Se-IV transformed to Se-V at a pressure around 80 GPa. Rietveld refinement of the Se-V phase at 110(3) GPa (Fig. 2d) on the basis of $\beta$-Po structure gives the following lattice parameters: $a_h$ = 3.8411(5) Å, $c_h$ = 2.8614(2) Å (in hexagonal setting), and $a_r$ = 2.414(1) Å, $\alpha_r$ = 105.42(1)$^o$ (in rhombohedral setting), with atomic volume V$at$ =

12.18(1) Å³. We observe the Se-V phase to be stable up to 140(3) GPa, where it transforms to Se-VI phase with *bcc* structure, in accordance with Ref. 12. The relationship between the $\beta$-Po and *bcc* structures is shown in Fig. 3(c).

The pressure dependence of the lattice parameters of the *bcm* and $\beta$-Po phases in S and Se is shown in Fig. 4 (upper and middle panels). The relation between the *bcm* structure and the primitive rhombohedral $\beta$-Po structure can be seen in Fig. 3(b), where $a_m = a_r (3 + 6cos\alpha_r)^{1/2} = c_h$, $b_m = a_r(2 - 2cos\alpha_r)^{1/2} = a_h$, $c_m = a_r$, and $\alpha_m = arccos(-(1/3 + 2/3cos\alpha_r)^{1/2})$. The transition from the modulated *bcm* to $\beta$-Po phase is nearly second order for both S and Se.

The structural data obtained in the present work in combination with the data available in the literature allow us to analyze the interatomic distances of Se over a wide pressure range from ambient pressure up to 150 GPa, through the transition from semiconducting to metallic state. Apart from Se phases II and III,[33,34] the data on the interatomic distances of all the known phases of Se are summarized in Figure 5. These are Se-I (trigonal structure with triangular atomic chains),[32] Se-VII (tetragonal structure with squared atomic chains),[28] Se-IV (modulated *bcm*) (Ref. 7 and present work) and Se-V (rhombohedral $\beta$-Po) (Ref. 12 and present work). Se-VII can be obtained at room temperature after heating Se-III to 450 K, and also known to transform to Se-IV above 28 GPa on pressure increase.[28] The nearest interatomic distance of the chain structures Se-I and Se-VII (2.36 Å) is determined by covalent bonding and is pressure independent. The second nearest interatomic distance (4 distances) of Se-I and Se-VII corresponds to the interchain distance and decreases drastically with pressure. In the metallic Se-IV phase with a modulated *bcm* structure, a new interatomic distance of 2.6 Å appears, indicating a change in the atomic bonding in comparison to Se-VII. The distance of 2.6 Å is unaffected by modulation and connects the atoms in linear chains running along the *c*-axis of the *bcm* cell of Se-IV. In addition, in Se-IV there are 2+2 distances that are modulated and their values are ranging between min and max values shown in Fig. 5. The minimum value of the modulated distance in Se-IV is limited by the covalent bonding. At higher pressure, in the $\beta$-Po Se-V structure, the 2+2+2 distances become equal and decrease slightly with pressure. The ultimate high-pressure phase of Se, Se-VI, has 8+6 nearest neighbors in its *bcc* structure. A very similar picture of interatomic distances is expected for sulfur in its nonmetallic phases II and III (triangular and squared chain structures),[28] modulated *bcm* structure of phase IV and $\beta$-Po structure of phase V.

## IV. Discussion

Superconducting transition temperature in Te and Se in the pressure range corresponding to the modulated *bcm* and $\beta$-Po phases show strongly non-linear behavior with a maximum T*c* = 4.2 K at 5.3 GPa in Te and 5.8 K at 25 GPa in Se with a decrease of T*c* on further compression[25,35,36] (for T*c* in Se, see Fig. 3, lower panel). Theoretical calculations of T*c* for Se and Te in this pressure range, using the $\beta$-Po structure, reproduce the value of T*c* and its decrease with increasing pressure.[37,38] A jump to a higher T*c* of 7.4 K is observed in Te at higher pressure of 35 GPa[36] (probably corresponding to a phase transition to *bcc*, Ref. 11), confirmed theoretically.[37] For *bcc* Se, T*c* is calculated as 10-11 K at 120-130 GPa with a decrease in T*c* with further pressure increase,[38,39] which agrees with the experimental data at 150-170 GPa.[25]

Metallic sulfur is superconducting with T$c$ of 10 K at 93 GPa, and the superconducting transition temperature rises up to 14 K at 157 GPa,[25,40] in contrast to a negative dT$c$/dP observed in the heavier superconducting Se and Te.[36] A jump in T$c$ up to 17 K is observed in S at 157 GPa, which is attributed to the *bco* - β-Po phase transition (Fig. 4, lower panel). The T$c$ for β-Po phase of sulfur is calculated as 17 K at 160 GPa[38,41] in agreement with the experiment. For a predicted *bcc* phase in S at 585 GPa, T$c$ is calculated as 15 K[21]. Another theoretical work that shows a simple cubic phase to be stable in S between β-Po and *bcc* phases, predicts T$c$ below 10 K at pressure of 280 GPa[38,41].

One can assume that the differences in the pressure dependence of T$c$ and electronic structure reported for S and Se[38] are connected with the difference in the pressure dependence of structural parameters for the modulated *bcm* phases. In Fig. 4, the pressure dependence of T$c$ in S and Se (lower panel) is shown in comparison with the pressure dependence of the structural parameters of the modulated *bcm* and β-Po phases. The difference in behavior of T$c$ correlates with the different behavior of both modulated vector $q$ and the cell angle of the *bcm* structure. The modulated vector $q$ in S decreases linearly with pressure, in contrast to the strong non-linearity of $q$ observed in Se.[7] Also, the pressure dependence of the cell angle β of the modulated *bcm* structures in S-IV shows a different behaviour if compared to that in Se-IV.

Calculations would be helpful to examine if the differences in structural behavior of the metallic phases of S and Se are connected with the electronic structure and differences in the pressure dependence of T$c$ in these phases.

## V. Conclusion.

We have shown that phase IV of sulfur stable above 83 GPa is incommensurately modulated, as found earlier for Se-IV and Te-III. The modulated *bcm* structure of S-IV and Se-IV is shown to transform gradually to the β-Po structure of S-V and Se-V at pressures of 153 GPa and approximately 80 GPa, respectively. Analysis of the interatomic distances in Se in the pressure range from ambient to 150 GPa shows the changes in bonding properties on compression from covalent to metallic. Pressure dependence of structural parameters in the modulated *bcm* and β-Po phases is found to be essentially different for S and Se, which may account for the difference in the pressure dependence of superconducting transition temperatures in these materials.

## Acknowledgements

This work and HPCAT is supported by DOE-BES, DOE-NNSA, DOD-TACOM, NSF, NASA, and the W.M. Keck Foundation. The authors acknowledge financial support from NSF, through grant EAR-0217389. The Advanced Photon Source is supported by the U. S. Department of Energy, Office of Science, Office of Basic Energy Sciences, under Contract No. W-31-109-Eng- 38.

**Figures**

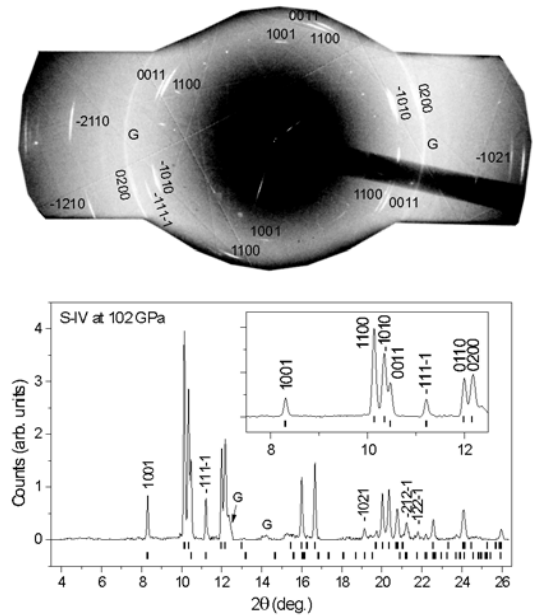

FIG. 1: (Upper panel) A diffraction image from a poor single crystal of S-IV at 102(3) GPa. Some of the diffraction reflections are indexed with their *hklm* indices. A powder line from the tungsten gasket is marked "G". (Lower panel) Diffraction profile of incommensurate S-IV at 102 GPa, obtained from the integration of the above image, collected with $\lambda = 0.3675$ Å. The upper and lower tick marks below the profile show the peak positions of the main and first-order satellite reflections, respectively. The inset shows some of the low-angle reflections. Some of the diffraction reflections are indexed with their *hklm* indices. A diffraction peak from gasket is marked "G".

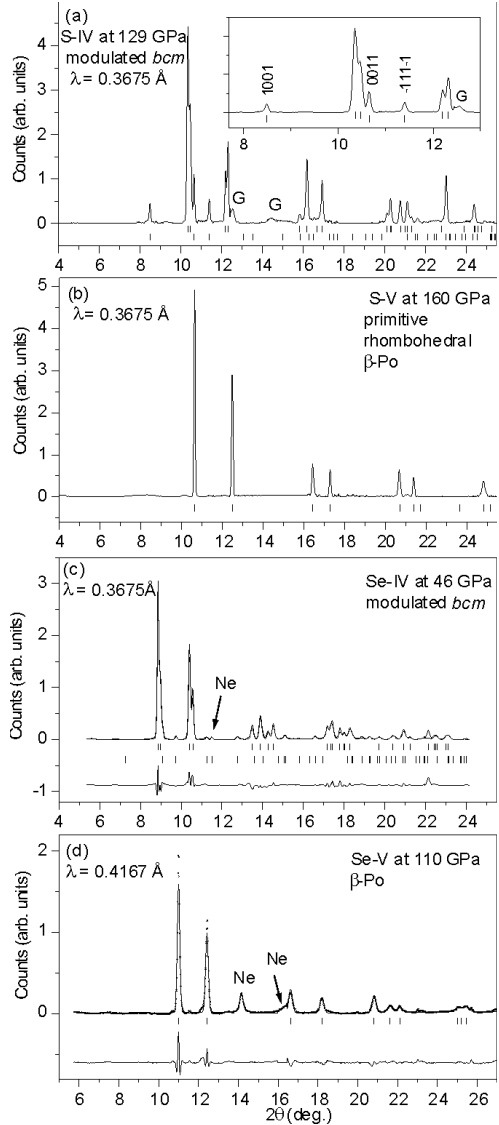

FIG. 2: Integrated x-ray diffraction spectra of (a) S-IV at 102(3) GPa, (b) S-V at 160(3) GPa, (c) Se-IV at 46(1) GPa and (d) Se-V at 110 (3) GPa, taken on pressure increase. Full profile Rietveld refinement is shown for Se-IV on the basis of the modulated *bcm* structure and for S-V on the basis of the $\beta$-Po structure. Dots and lines denote the observed and calculated profiles, respectively. The tick marks below all profiles indicate the predicted peak positions. For S-IV and Se-IV, the upper and lower tick marks correspond to the positions of the main and modulation reflections, respectively. Below the tick marks is the difference curve between the observed and calculated profiles. "G" denotes the diffraction reflections from gasket material (tungsten), and "Ne" indicates the diffraction peaks from neon used as pressure transmitting medium.

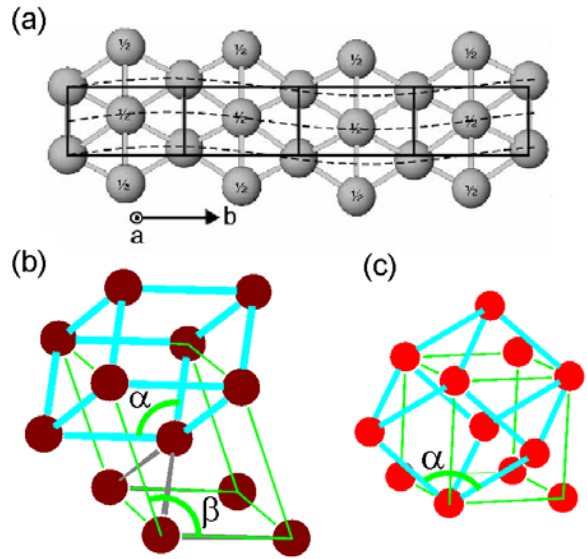

FIG. 3: Structural relationship of the metallic phases in the group-VI elements. (a) The modulated *bcm* structure of Te-III and Se-IV (four unit cells) viewed down the a-axis (from Ref. 6). The six nearest-neighbor distances for each atom are shown, and those atoms at the body centers are labeled with "1/2". (b) Structural relation between the *bcm* structure of S-IV (thin lines) and a primitive rhombohedral structure of *β*-Po-type (thick lines). The monoclinic angle *β* and the rhombohedral angle *α* are shown. (c) Structural relation between the primitive rhombohedral structure (thick lines) and the body-centered cubic structure (thin lines) in Se. The rhombohedral angle *α* (=109.47*o*) is indicated.

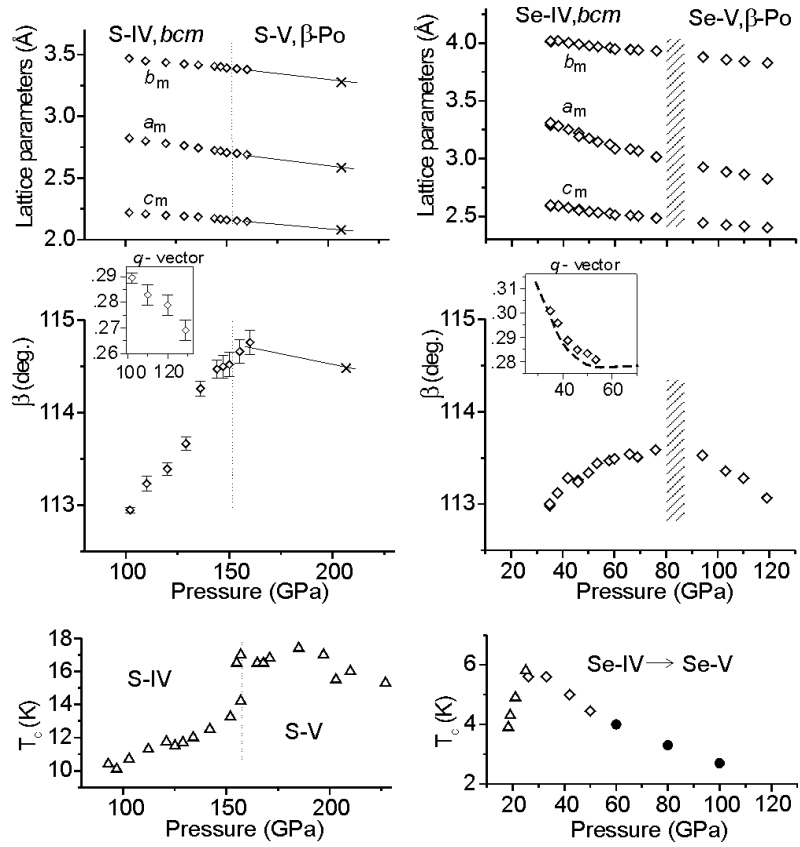

FIG. 4: (Upper and middle panel) Pressure dependence of lattice parameters for the body-centered monoclinic (*bcm*) S-IV and Se-IV and rhombohedral $\beta$-Po S-V and Se-V structures (in monoclinic setting). Open diamonds denote data from the present work. The data at 206 GPa for S-V denoted with "x" is taken from Ref. 20. Insets in the middle panel show the pressure dependence of the modulation vector $q$, with open diamonds denoting data from present work, and the dashed line denoting data from Ref. 7. Solid lines are guides to the eye. (Lower panel) Pressure dependence of the temperature of superconducting transition in S and Se, with open symbols denoting experimental data (triangles from Ref. 25, diamonds from Ref. 36) and solid symbols denoting data from theoretical work.[38] Vertical dotted lines and dashed areas denote proposed transition pressures.

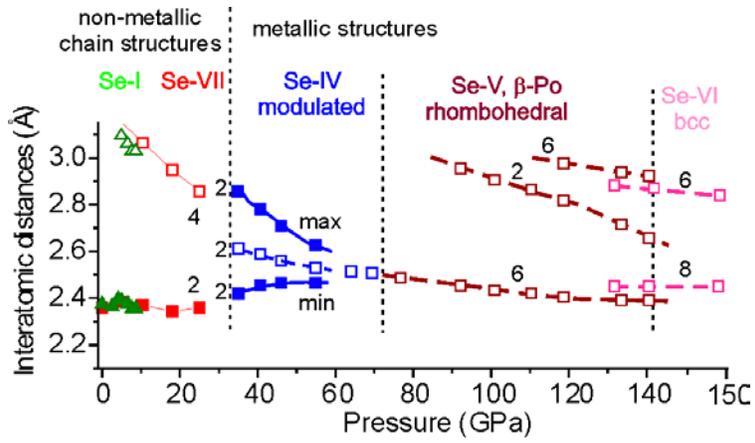

FIG. 5: Interatomic distances for the Se phases up to 150 GPa. The data on Se-I (green triangles) are taken from Ref. 32, the data on Se-VII (red squares) are taken from Ref. 28. The data on Se-IV, V and VI (blue, brown and pink squares) are from present work and are in agreement with the literature data.[7,12] The numbers indicate how many of the distances are present in the structure. The vertical dotted lines indicate the proposed transition pressures.